\documentclass[journal]{IEEEtran}

\usepackage{xcolor,soul,framed} 
\usepackage{makecell}

\usepackage[colorlinks=true,allcolors=black]{hyperref}
\colorlet{shadecolor}{yellow}
\usepackage[pdftex]{graphicx}
\graphicspath{{../pdf/}{../jpeg/}}
\DeclareGraphicsExtensions{.pdf,.jpeg,.png}
\usepackage{array}
\usepackage[cmex10]{amsmath}
\usepackage{mdwmath}
\usepackage{mdwtab}
\usepackage{eqparbox}
\usepackage{url}
\usepackage{graphicx}
\usepackage{pgfplots}
\usepgfplotslibrary{groupplots}
\usepackage{pgf-pie}
\usepackage{tabularx}
\usepackage{makecell}
\usepackage{multirow}
\usepackage{pgfgantt}
\usepackage{float} 
\usepackage[english]{babel}
\usepackage[utf8]{inputenc}
\usepackage{csquotes}
\usepackage[style=numeric,sorting=none, backend=biber]{biblatex}
\usepackage{titlesec}
\usepackage{stfloats}

\begin{document}
\bstctlcite{IEEEexample:BSTcontrol}
    \title{AI-powered Contextual 3D Environment Generation: A Systematic Review\thanks{\scriptsize This work has been submitted to the IEEE for possible publication. Copyright may be transferred without notice, after which this version may no longer be accessible.}}

\author{
    \IEEEauthorblockN{Miguel Silva\IEEEauthorrefmark{1}, Alexandre Valle de Carvalho\IEEEauthorrefmark{1}\IEEEauthorrefmark{2}}
    \IEEEauthorblockA{\IEEEauthorrefmark{1}Faculty of Engineering, University of Porto, Portugal}
    \IEEEauthorblockA{\IEEEauthorrefmark{2}INESC TEC, Portugal}
}


\maketitle
\begin{abstract}
The generation of high-quality 3D environments is crucial for industries such as gaming, virtual reality, and cinema, yet remains resource-intensive due to the reliance on manual processes. This study performs a systematic review of existing generative AI techniques for 3D scene generation, analyzing their characteristics, strengths, limitations, and potential for improvement. By examining state-of-the-art approaches, it presents key challenges such as scene authenticity and the influence of textual inputs. Special attention is given to how AI can blend different stylistic domains while maintaining coherence, the impact of training data on output quality, and the limitations of current models. In addition, this review surveys existing evaluation metrics for assessing realism and explores how industry professionals incorporate AI into their workflows. The findings of this study aim to provide a comprehensive understanding of the current landscape and serve as a foundation for future research on AI-driven 3D content generation. Key findings include that advanced generative architectures enable high-quality 3D content creation at a high computational cost, effective multi-modal integration techniques like cross-attention and latent space alignment facilitate text-to-3D tasks, and the quality and diversity of training data combined with comprehensive evaluation metrics are critical to achieving scalable, robust 3D scene generation.
\end{abstract}


%
\IEEEpeerreviewmaketitle


\section{Introduction}

The creation of complex 3D scenes is essential for industries such as gaming, virtual reality, and cinema, but remains a resource intensive process due to the dependence on manual labor and expertise \cite{smelik2014survey, samavati2023deep}. As demand for immersive high-quality environments grows, current procedural generation tools face significant limitations, including high production costs, time-consuming workflows, and a lack of realism and intuitive control \cite{smelik2014survey}. These constraints hinder creative freedom, forcing artists and developers to work within rigid frameworks, ultimately reducing efficiency in content creation.

This work performs a systematic review of state-of-the-art generative AI (Gen AI) approaches for 3D scene generation, analyzing their effectiveness in creating large, contextually coherent environments. By examining automation techniques, contextual integration, and the role of textual prompts, this study aims to provide a comprehensive understanding of AI-driven methodologies developed in the last four years, a period marked by the rapid evolution of Gen AI to highlight its strengths and limitations without focusing on previous foundational work \cite{summerville2018procedural, hong20233dllminjecting3dworld}.

This work was performed within a broader scope, targeting research on contextually informed layout generation to iteratively expand large-scale, coherent 3D environments.

\subsection{Challenges in Generative 3D Scene Creation}

Despite advances in procedural content generation, current AI-driven approaches still struggle with key challenges, including realism, context awareness, and adaptability. Existing terrain and mesh generation tools often rely on procedural algorithms, leading to repetitive textures and patterns that diminish immersion \cite{smelik2014survey, samavati2023deep}. In addition, this research did not find standard approaches for seamlessly integrating newly generated assets with preexisting environments, making it difficult to achieve natural transitions between distinct regions.

A crucial aspect of improving Gen AI for the generation of 3D content is enhancing its ability to infer and maintain contextual consistency throughout the generation process. Furthermore, textual prompts are an emerging method for guiding AI-driven content creation, yet their effectiveness in influencing the composition and structure of the scene remains an open research question \cite{höllein2023text2roomextractingtextured3d, shriram2024realmdreamertextdriven3dscene}. 

This work, consisting of a literature review and critical analysis, addresses how current models handle these challenges and explores potential improvements to Gen AI techniques to bridge the gaps in automation, coherence, and creative flexibility.

\section{Methodology} \label{chap:literature_review}

\begin{figure*}[h]
    \centering 
    \includegraphics[width=0.8\linewidth]{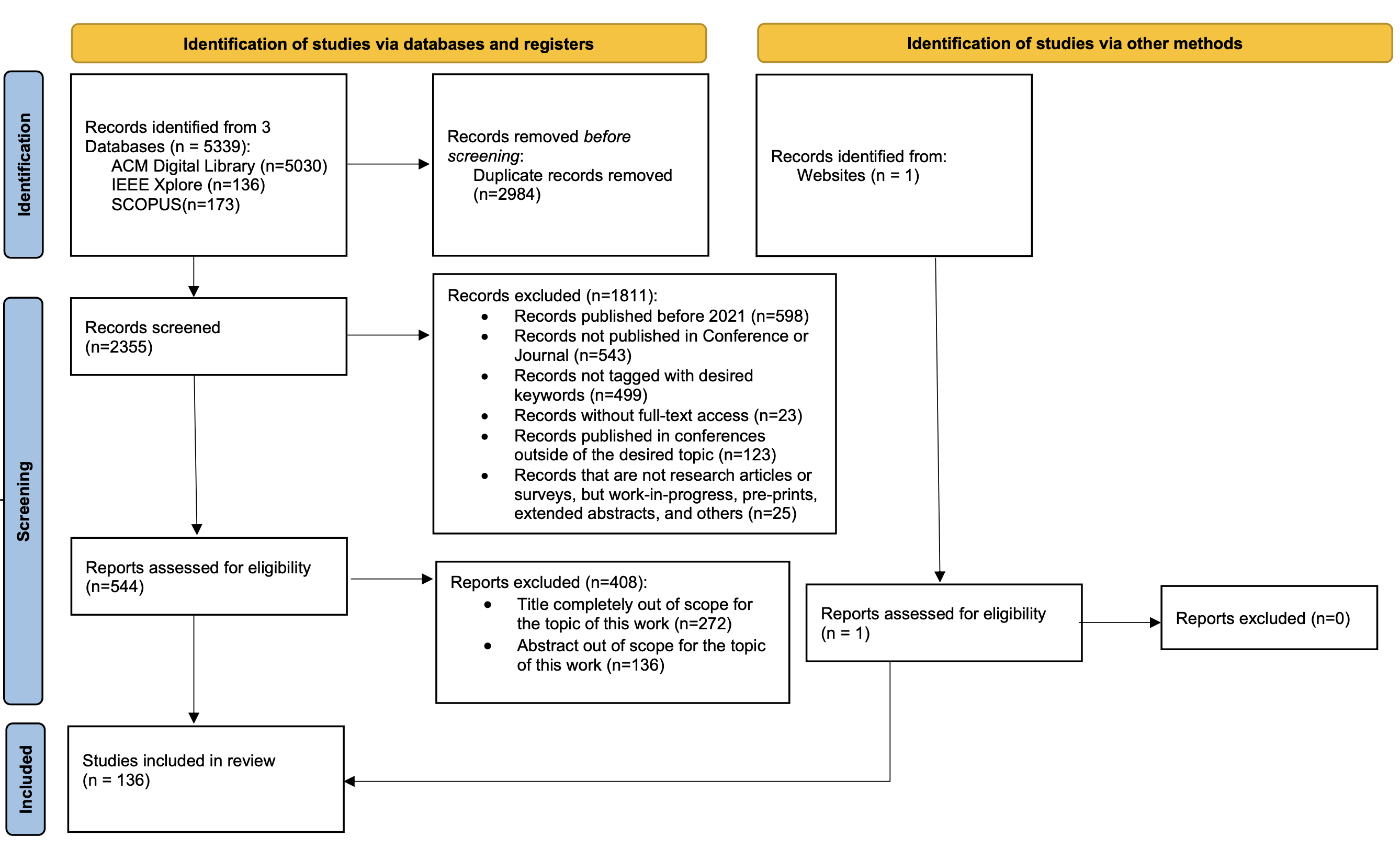}
    \caption{PRISMA Flowchart}
    \label{fig:prisma-flow}
\end{figure*}

The literature review was performed using Preferred Reporting Items for Systematic Reviews and Meta-analysis (PRISMA \cite{prisma}). Zotero was used to manage, categorize, and store references, while Python supported data processing through the use of the Pandas library. Figure \ref{fig:prisma-flow} summarizes the process with the results obtained from the application of the PRISMA method.

\subsection{Data Sources}
\label{sec:databases}

To initially populate the first stage of PRISMA with records, three databases were considered: IEEE, ACM, and Scopus. Popular indexers and other publication sites, such as Google Scholar and \textit{arXiv.org}, were excluded due to the heterogeneity of the results in the publication mediums, which present a lot of preprints and work that were not peer-reviewed to the current date.

Other records were included (as allowed by the PRISMA framework) resulting from other sources and as part of the exploratory work, which in this case resulted in the review of one additional record \cite{nvidia2024edify3dscalablehighquality}.

\subsection{Search Queries}
\label{sec:searchqueries}

The search queries are constructed iteratively, with a strong emphasis on refining and expanding the scope of results to comprehensively capture the applications of Gen AI in 3D synthesis. However, records on this topic for application domains such as gaming, cinema, and robotics were also targeted.
\subsubsection{Refinement Process}

After an initial draft was produced, iterative testing of the search queries was performed to decide on the definitive version. The most notable changes that are still present in the final version are the following.

\begin{itemize}
    \item \textbf{Removal of common string parts}: The term  \textit{(generation OR creation OR modeling OR synthesis)} was part of all queries before removing it, as the results showed little to no difference in having it. The term \textit{generative}, which was initially included in all queries, semantically overlaps with the meaning that was conveyed in the removed term, hence the lack of distinct results.
    \item \textbf{Improving on the term \textit{AI}}: The term \textit{AI} was not so popular in an academic setting, leading to the inclusion of alternatives to its short-form in the final queries to combat its lack of popularity in searchable records.
    \item \textbf{Removal of the term \textit{World}}: this term introduced noise in the search, with synonyms like \textit{scene} or \textit{environment} performing much better.
    \item \textbf{Merging two queries}: There was a query dedicated to finding records related to the cross-modal applications of 3D Mesh generation models. In the end, the keyword \textit{Cross-Modal} was included in the existing search query that extracted information related to other applications and aspects of training for machine learning models.
\end{itemize}

The final version ended up comprising the following set of search queries:

 \begin{itemize}
     \item \textbf{Q1}. ("generative AI” OR “generative artificial intelligence" OR "generative model") AND 3D AND (scene OR content OR terrain OR environment) AND (procedural OR automated)
     \item \textbf{Q2}. ("generative artificial intelligence" OR "generative model" OR "generative AI”) AND (scene OR content OR terrain OR environment) AND 3d AND (real-time OR dynamic)
     \item \textbf{Q3}. ("generative artificial intelligence" OR "generative model" OR "generative AI”) AND ("few-shot learning" OR "domain adaptation" OR "transfer learning" OR "cross-modal”) AND 3d AND text OR image AND NOT video
     \item \textbf{Q4}. ("generative artificial intelligence" OR "generative model" OR "generative AI”) AND (gaming OR "virtual reality" OR cinema) AND 3d
 \end{itemize}

A thesaurus of keywords or sets of words used atomically in the context of querying is compiled to structure the search strategy, as summarized in the following table \ref{tab:thesaurus}.

{
\renewcommand{\arraystretch}{1.5} 

\begin{table}[h]
    \footnotesize
    \centering

    \begin{tabular}{|p{3.6cm}|p{2.1cm}|p{0.9cm}|}
        \hline
        \multicolumn{1}{|c|}{\textbf{Domain}} & 
        \multicolumn{1}{c|}{\textbf{Focus}} & 
        \multicolumn{1}{c|}{\textbf{Constraints}} \\ \hline
        \multirow{4}{3cm}{\raggedright\makecell[l]{Generative AI \\ Generative Artificial Intelligence \\ Generative Model \\ 3D}} 
        & \multirow{2}{2cm}{\makecell[c]{\\[1pt] Scene \\ Content \\ Terrain \\ Environment }}  
          & \makecell[l]{\\[1pt] Real-time \\ Dynamic}\\ \cline{3-3}
        & & \makecell[l]{\\[1pt] Procedural \\ Automated} \\ \cline{2-3}
        & \makecell[l]{\\Few-Shot Learning \\Domain Adaptation \\Transfer Learning \\ Cross-modal} 
          & \makecell[l]{Text \\ Video \\ Image} \\ \cline{2-3} 
        & \makecell[c]{\\[1pt] Gaming \\ Virtual Reality \\ Cinema \\ [3pt]} 
          & - \\ 
        \hline
    \end{tabular}
    \caption{Search Keywords}
    \label{tab:thesaurus}
\end{table}
}
The keywords are categorized into three components: Domain of research (e.g., "generative AI"), Focus (target application or aspect of such domain, e.g., "scene generation," "few-shot learning," "gaming"), and Constraints (e.g., "real-time," "text-to-3D"). These terms are combined to form queries (Q1–Q4) to target specific research intersections. For instance, Q1 emphasizes procedural automation for 3D environments, while Q3 integrates cross-modal adaptation while explicitly excluding video-related works. Constraints such as "real-time" or "dynamic" are applied to prioritize practical applications, whereas keywords like "gaming" and "virtual reality" narrow the scope to entertainment and simulated scenarios. 

This structured approach ensures comprehensive coverage of generative AI’s role in 3D content creation, balancing specificity with breadth across modalities, learning paradigms, and application contexts.

\subsection{Eligibility}
\label{sec:eligibility}

Records resulting from the search are considered  eligible if they further comply with the following criteria:

\begin{enumerate}
    \item Is not  duplicated nor retracted in the final collection.
    \item Provide full-text access.
    \item Published in the year 2021 or after.
    \item Published in the English Language.
    \item Published in a journal or a conference.
    \item Published under topics related to this work's theme of research, focusing on areas strictly related to Artificial Intelligence, Multimedia, or, more generically, Computer Science.
    \item Automatically tagged with the desired keywords for the topic.
    \item Not a work-in-progress, a pre-print, or any other type of document different from a research article or survey.
\end{enumerate}

A total of 5340 records were identified, with 5339 retrieved from three databases (ACM Digital Library, IEEE Xplore, and SCOPUS) and 1 from a website. After removing 2984 duplicate records, 2355 records remained for screening. The screening against the selected criteria resulted in 1811 records being excluded. For each of the remaining 544 records, the title and abstract were read and analyzed to decide whether or not to include them in the process. This led to an additional 408 records being excluded. Finally, a full read was performed on the 136 records still considered relevant, which were then included in the review.

\subsubsection{Challenges when Excluding Records}

The database queries rendered a noticeable amount of records misidentified as relevant or related to the topic of research: that was because these records mention 3D technologies and generative models tangentially in their full text, even though those were not the main focus of research for such articles. One solution that greatly improved the quality of the included records for review was the addition of eligibility criteria \textit{Records must be automatically tagged with the desired keywords for the topic}. Using the reference manager's automatic record tagging feature, a few keywords were selected as crucial and something that all records should address to filter out records that would not be tagged with any of these words. The selected keywords were \textit{Generative}, \textit{GAN}, \textit{Graphics}, \textit{Model}, \textit{AI}, \textit{Mesh}, \textit{3D}, \textit{Autoencoder} and \textit{Attention}. Furthermore, addressing the area of publication of each record proved to be very useful as well, enabling the removal of articles that were more focused on the applications of emerging technologies rather than the technical aspects of their implementations.

\subsubsection{Manual Removal and Automatic Exclusion}

\begin{figure}
    \centering
    \begin{tikzpicture}
        \begin{axis}[
            ybar,
            bar width=0.3cm, 
            ymin=0, ymax=80,
            width=0.9\columnwidth, 
            height=4cm, 
            enlarge x limits=0.2, 
            symbolic x coords={2021, 2022, 2023, 2024},
            xtick=data,
            xticklabel style={rotate=90, anchor=east, font=\small}, 
            ylabel={Number of Records},
            ylabel style={font=\small}, 
            xlabel={Year},
            xlabel style={font=\small}, 
            nodes near coords, 
            nodes near coords align={vertical}, 
            nodes near coords style={font=\tiny}, 
            tick label style={font=\small}, 
        ]
            \addplot[fill=gray] coordinates {(2024, 71) (2023, 39) (2021, 13) (2022, 13)};
        \end{axis}
    \end{tikzpicture}
    \caption{Distribution of Records by Year}
    \label{fig:records_distribution}
\end{figure}
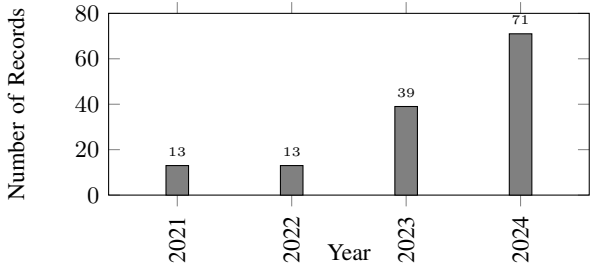

Some steps were performed manually due to the lack of automated tools. For example, the process of confirming that a paper is published in a conference or journal related to the topic of research, checking its properties to see if it was a work-in-progress or not, as well as validating the title and abstract of each record was all done manually. These checks helped exclude 556 of the 2355 unique records, while the remaining 1663 removed records were excluded with the use of automatic tools. The resulting distribution of included articles per database is found in figures \ref{fig:before_screening} and \ref{fig:after_screening}. Figure \ref{fig:records_distribution} highlights the distribution of included articles across the years, and table \ref{tab:query_dist} summarizes the contribution of each database per search query.

\begin{figure}
    \centering
    \begin{minipage}{0.45\textwidth}
        \centering
        \begin{tikzpicture}
            \pie[
                text=legend,
                radius=2,
                explode=0.1,
                color={white, gray, black!25} 
            ]{
                94.2/ACM,
                2.5/IEEE XPlore,
                3.2/SCOPUS
            }
        \end{tikzpicture}
        \caption{Distribution of Records before Screening}
        \label{fig:before_screening}
    \end{minipage}
    \hfill
    \begin{minipage}{0.45\textwidth}
        \centering
        \begin{tikzpicture}
            \pie[
                text=legend,
                radius=2,
                explode=0.1,
                color={white, gray, black!25} 
            ]{
                76.3/ACM,
                10.4/IEEE XPlore,
                13.3/SCOPUS
            }
        \end{tikzpicture}
        \caption{Distribution of Records After Screening}
        \label{fig:after_screening}

    \end{minipage}
\end{figure}

\begin{table}[h]
    \centering
    \begin{tabular}{|l|r|r|r|r|r|}
        \hline
        & \textbf{Q1} & \textbf{Q2} & \textbf{Q3} & \textbf{Q4} & \textbf{Total} \\
        \hline
        \textbf{IEEE Xplore} & 3 & 8 & 3 & 0 & 14 \\
        \hline
        \textbf{SCOPUS} & 2 & 5 & 2 & 9 & 18\\
        \hline
        \textbf{ACM Digital Library} & 79 & 20 & 3 & 1 & 103\\
        \hline
    \end{tabular}
    \caption{Included Articles per Query and Database}
    \label{tab:query_dist}
\end{table}

\subsection{Annotation}
\label{sec:annotation}

The final collection of records is obtained from the search records and after completing the eligibility process, with the documents being fetched using Zotero's "Find Full Text" tool. Each document was fully read and subject to an assessment questionnaire to answer the following questions.

\begin{itemize}
    \item What is the target industry or application domain?
    \item What problem is solved?
    \item Which AI architecture is implemented?
    \item What multi-modal techniques are used?
    \item What is the input?
    \item What is the output?
    \item What experimental method is used?
    \item What metrics are considered in the experimental method?
    \item Which datasets are used?
    \item What are the limitations, gaps, challenges, and future work?
\end{itemize}

Together, these questions ensure a structured and critical assessment of the technical foundations, empirical validity, and practical relevance of each work. They capture a) the core architectural and algorithmic contributions; b) the soundness of experimental validation and transparency regarding constraints; c) the work within its applied context, assessing its relevance and potential. The assessment helped to achieve a more clear understanding of best practices and limitations within the field and highlight aspects that are of critical importance in future research.

The following sections structure the findings from the perspective of a literature review and critical analysis, by thematic focus.

\section{Review and Analysis}

First, the primary challenges addressed in the literature are outlined, followed by an analysis of the architectures proposed to generate 3D content. Subsequent sections examine techniques to ensure aesthetic coherence, identify current methodological limitations, and conclude with strategies for integrating multiple input modalities in generative models, as well as highlighting their applications in various domains and the evaluation metrics used to assess model performance.

\subsection{Problems Addressed}

The state-of-the-art research in 3D content creation highlights a clear trend toward integrating AI-driven automation, realism, and adaptability across tasks. Key advancements focus on bridging 2D and 3D domains, with methods for generating 3D assets from images \cite{liu_3dall-e_2023}, synthesizing novel views, and enabling interactive design. Semantic understanding through segmentation and procedural generation further enhances realism and efficiency, while style transfer in 2D and 3D enables creative flexibility. Automation in animation, rigging, and object placement simplifies workflows, while dataset augmentation and anomaly detection ensure robustness and scalability.

The works were categorized according to the specific problems and topics they addressed, as outlined in table \ref{tab:task_references}, facilitating a structured analysis of the relevant literature.

\begin{table*}[t]
\centering
\renewcommand{\arraystretch}{1.3} 
\small 
\begin{tabularx}{\textwidth}{X X}
\hline
Task & References \\ 
\hline
Interactive 3D Design & \cite{liu_3dall-e_2023}, \cite{sha_nerf-is_2024}, \cite{zhao_vaide_2024}, \cite{hertz_spaghetti_2022}, \cite{hu_generative_2024}, \cite{li_sp-gan_2021}, \cite{wang_roomdreaming_2024}, \cite{he_creating_2024}, \cite{de_la_torre_llmr_2024}, \cite{kim_neuralfield-ldm_2023}, \cite{dong_coin3d_2024}, \cite{hu_cns-edit_2024}, \cite{kikuchi_constrained_2021}, \cite{gao_scenehgn_2023}, \cite{maesumi_explorable_2023}, \cite{li_icontrol3d_2024}, \cite{chen_memovis_2024}, \cite{wang_proteusnerf_2024}, \cite{zhang_protodreamer_2024} \\ 
Image Generation & \cite{liu_3dall-e_2023}, \cite{harris_multi-stage_2021}, \cite{rajaram_blendscape_2024}, \cite{jabbar_survey_2021}, \cite{rosenberg_drawtalking_2024}, \cite{huang_blue_2024}, \cite{alaluf_neural_2023}, \cite{chung_artinter_2023}, \cite{cao_autoencoder-based_2024}, \cite{chen_find_2024}, \cite{richardson_conceptlab_2024} \\ 
3D Scene Generation & \cite{li_sp-gan_2021}, \cite{wang_roomdreaming_2024}, \cite{he_creating_2024}, \cite{de_la_torre_llmr_2024}, \cite{kim_neuralfield-ldm_2023}, \cite{gao_scenehgn_2023}, \cite{maesumi_explorable_2023}, \cite{li_icontrol3d_2024}, \cite{yang_geolatent_2023}, \cite{li_sat2scene_2024}, \cite{jiang_synthesizing_2021}, \cite{zuo_dg3d_2023}, \cite{gothoskar_3dp3_2021}, \cite{mo_sparse_2024}, \cite{huang_plantography_2024}, \cite{yiyang_context-aware_2024}, \cite{gottsacker_rlty2rlty_2024}, \cite{han_scene_2024}, \cite{son_singraf_2023}, \cite{wang_rip-nerf_2023}, \cite{merino_interactive_2023}, \cite{awiszus_world-gan_2021}, \cite{petrov_gem3d_2024}, \cite{mohammad_khalid_clip-mesh_2022}, \cite{zhang_vrcopilot_2024}, \cite{dang_worldsmith_2023}, \cite{si_generating_2024}, \cite{shen_neural_2024}, \cite{wu_blockfusion_2024}, \cite{cai_l-magic_2024}, \cite{kant_invs_2023}, \cite{singh_worldgen_2023}, \cite{jones_shapecoder_2023}, \cite{rai_assessing_2024} \\ 
3D Reconstruction & \cite{sha_nerf-is_2024}, \cite{zhao_vaide_2024}, \cite{hu_generative_2024}, \cite{li_sp-gan_2021}, \cite{de_la_torre_llmr_2024}, \cite{dong_coin3d_2024}, \cite{gao_scenehgn_2023}, \cite{wang_proteusnerf_2024}, \cite{zhang_protodreamer_2024}, \cite{yang_geolatent_2023}, \cite{zuo_dg3d_2023}, \cite{gothoskar_3dp3_2021}, \cite{mo_sparse_2024}, \cite{son_singraf_2023}, \cite{wang_rip-nerf_2023}, \cite{petrov_gem3d_2024}, \cite{mohammad_khalid_clip-mesh_2022}, \cite{wu_blockfusion_2024}, \cite{cai_l-magic_2024}, \cite{kant_invs_2023}, \cite{chen_shaddr_2023}, \cite{chen_lart_2023}, \cite{wu_dreamup3d_2024}, \cite{zhang_clay_2024}, \cite{ma_multicad_2023}, \cite{zhou_unsupervised_2022}, \cite{ryu_360_2023}, \cite{cheng_cross-modal_2022}, \cite{kimura_chartpointflow_2021}, \cite{yan_online_2021}, \cite{wang_elevating_2023}, \cite{eckart_self-supervised_2021}, \cite{perla_easi-tex_2024}, \cite{sun_f-3dgs_2024}, \cite{zhang_3dshape2vecset_2023}, \cite{wu_hyperdreamer_2023}, \cite{zhang_end--end_2022}, \cite{de_sousa_ribeiro_object-centric_2024}, \cite{wu_learning_2022}, \cite{xiao_text-free_2024}, \cite{nakada_study_2023} \\ 
3D Segmentation & \cite{sha_nerf-is_2024}, \cite{kim_neuralfield-ldm_2023}, \cite{gothoskar_3dp3_2021}, \cite{zhang_vrcopilot_2024}, \cite{singh_worldgen_2023}, \cite{ma_multicad_2023}, \cite{kimura_chartpointflow_2021}, \cite{yan_online_2021}, \cite{eckart_self-supervised_2021}, \cite{zhang_3dshape2vecset_2023}, \cite{de_sousa_ribeiro_object-centric_2024}, \cite{wang_embracing_2021}, \cite{faruqi_style2fab_2023}, \cite{zhou_edit3d_2024}, \cite{guerrero-viu_texsliders_2024}, \cite{sella_spicee_2024} \\ 
Text-to-3D Generation & \cite{he_creating_2024}, \cite{huang_plantography_2024}, \cite{yiyang_context-aware_2024}, \cite{mohammad_khalid_clip-mesh_2022}, \cite{si_generating_2024}, \cite{zhang_clay_2024}, \cite{zhang_3dshape2vecset_2023}, \cite{zhang_text--3d_2023}, \cite{ye_relscene_2024}, \cite{li_cad_2024}, \cite{li_imagebind3d_2024}, \cite{zhong_dreamlcm_2024}, \cite{weng_dream_2024}, \cite{huang_placiddreamer_2024}, \cite{yin_text2vrscene_2024}, \cite{polys_prompt_2024}, \cite{zheng_sketch3d_2024} \\ 
Image-to-3D Conversion & \cite{zhang_clay_2024}, \cite{zhang_3dshape2vecset_2023}, \cite{li_imagebind3d_2024}, \cite{numan_spaceblender_2024} \\ 
Dataset Augmentation & \cite{eckart_self-supervised_2021}, \cite{zhang_3dshape2vecset_2023}, \cite{de_sousa_ribeiro_object-centric_2024}, \cite{lee_knowledge_2023} \\ 
Novel View Synthesis & \cite{harris_multi-stage_2021}, \cite{alaluf_neural_2023}, \cite{ryu_360_2023}, \cite{zhou_edit3d_2024} \\ 
Style Transfer in 2D & \cite{jabbar_survey_2021}, \cite{chung_artinter_2023}, \cite{chen_controlstyle_2023} \\
Style Transfer in 3D & \cite{wang_roomdreaming_2024}, \cite{chen_shaddr_2023}, \cite{zhang_clay_2024}, \cite{perla_easi-tex_2024}, \cite{faruqi_style2fab_2023}, \cite{guerrero-viu_texsliders_2024}, \cite{sella_spicee_2024}, \cite{vecchio_controlmat_2024}, \cite{hu_diffusion_2024}, \cite{guerrero_matformer_2022}, \cite{zhang_mapa_2024}, \cite{zhou_photomat_2023} \\ 
Design Optimization & \cite{hertz_spaghetti_2022}, \cite{hu_cns-edit_2024}, \cite{gao_scenehgn_2023}, \cite{maesumi_explorable_2023}, \cite{zhang_protodreamer_2024}, \cite{jiang_synthesizing_2021}, \cite{merino_interactive_2023}, \cite{jones_shapecoder_2023}, \cite{polys_prompt_2024}, \cite{zhou_photomat_2023}, \cite{rios_large_2023} \\ 
Graph Generation & \cite{upadhyay_floorgan_2023} \\ 
Procedural Generation & \cite{hu_generative_2024}, \cite{wang_roomdreaming_2024}, \cite{gao_scenehgn_2023}, \cite{wang_proteusnerf_2024}, \cite{merino_interactive_2023}, \cite{awiszus_world-gan_2021}, \cite{shen_neural_2024}, \cite{wu_learning_2022}, \cite{guerrero-viu_texsliders_2024}, \cite{yin_text2vrscene_2024} \\
Animation and Rigging & \cite{yin_text2vrscene_2024}, \cite{polys_prompt_2024} \\ 
\hline
\end{tabularx}
\caption{Works classified by the problem they attempt to solve}
\label{tab:task_references}
\end{table*}

Addressing scene generation, recent advancements highlight the transformative role of Gen AI in this domain. 3DP3: 3D Scene Perception via Probabilistic Programming \cite{gothoskar_3dp3_2021} demonstrates the potential of probabilistic programming to model and interpret large terrains, offering robust scene understanding through probabilistic modulation. BlockFusion \cite{wu_blockfusion_2024} architecture, on the other hand, uses latent tri-plane extrapolation\footnote{Representing a 3D volume using three orthogonal 2D feature planes (XY, YZ, XZ). This allows efficient processing by projecting 3D data into 2D slices and extrapolating features.} and Signed Distance Functions (SDFs)\footnote{ Scalar field where each point in space stores the shortest distance to the surface of an object, with the sign indicating whether the point is inside (negative) or outside (positive) the object.} to compress 3D scenes effectively, providing a scalable approach to scene generation expanding one block at a time. It successfully introduces a new paradigm of generating 3D scenes by producing cohesive blocks of the environment at a time, seamlessly integrating them into the scene without explicitly placing or arranging previously generated elements. Other solutions, however, rely heavily on pre-defined layouts to position individual assets within the environment. Such is the case of Edify3D \cite{nvidia2024edify3dscalablehighquality}, which obtained remarkable results with the latter approach, leveraging a multi-view diffusion model to generate said assets. 

\begin{figure}
    \centering
    \includegraphics[width=1\linewidth]{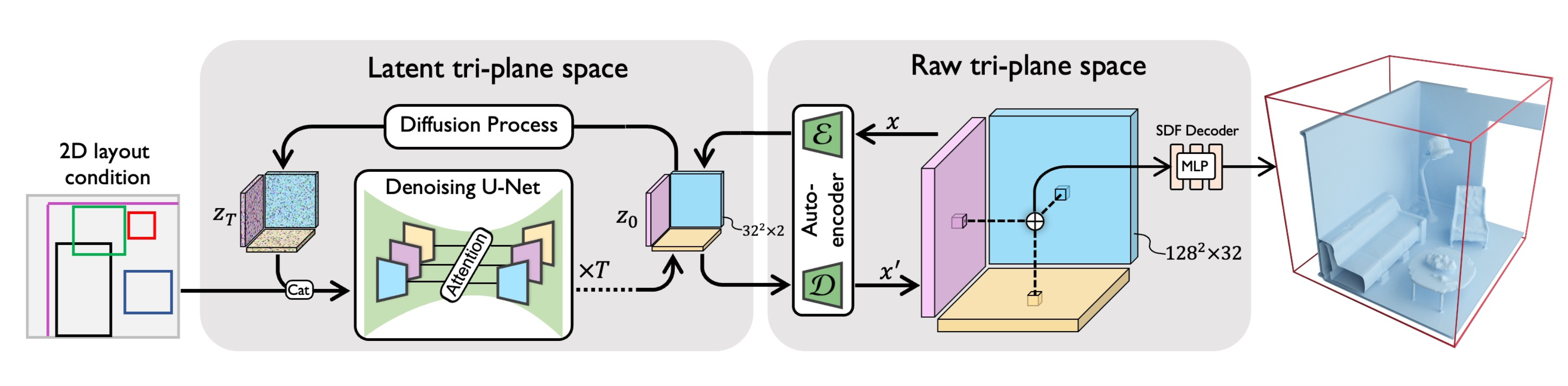}
    \caption{BlockFusion generates new blocks given 2D Conditional Layout.}
    \label{fig:blockfusion-triplane}
\end{figure}

Beyond cohesive scene generation, some works explore advanced synthesis and editing mechanisms. DG3D \cite{zuo_dg3d_2023} combines GANs with multimodal diffusion renderings to produce high-quality 3D textured shapes, merging two powerful generative paradigms. Explorable Mesh Deformation Subspaces from Unstructured 3D Generative Models \cite{maesumi_explorable_2023} employ k-nearest-neighbor techniques for intuitive and granular shape deformations, while GEM3D \cite{petrov_gem3d_2024} breaks down the diffusion process into medial abstractions, enabling detailed skeletal synthesis for 3D shapes. These advances emphasize the importance of structural abstraction in creating functional assets.

\begin{figure}[H]
    \centering
    \includegraphics[width=1\linewidth]{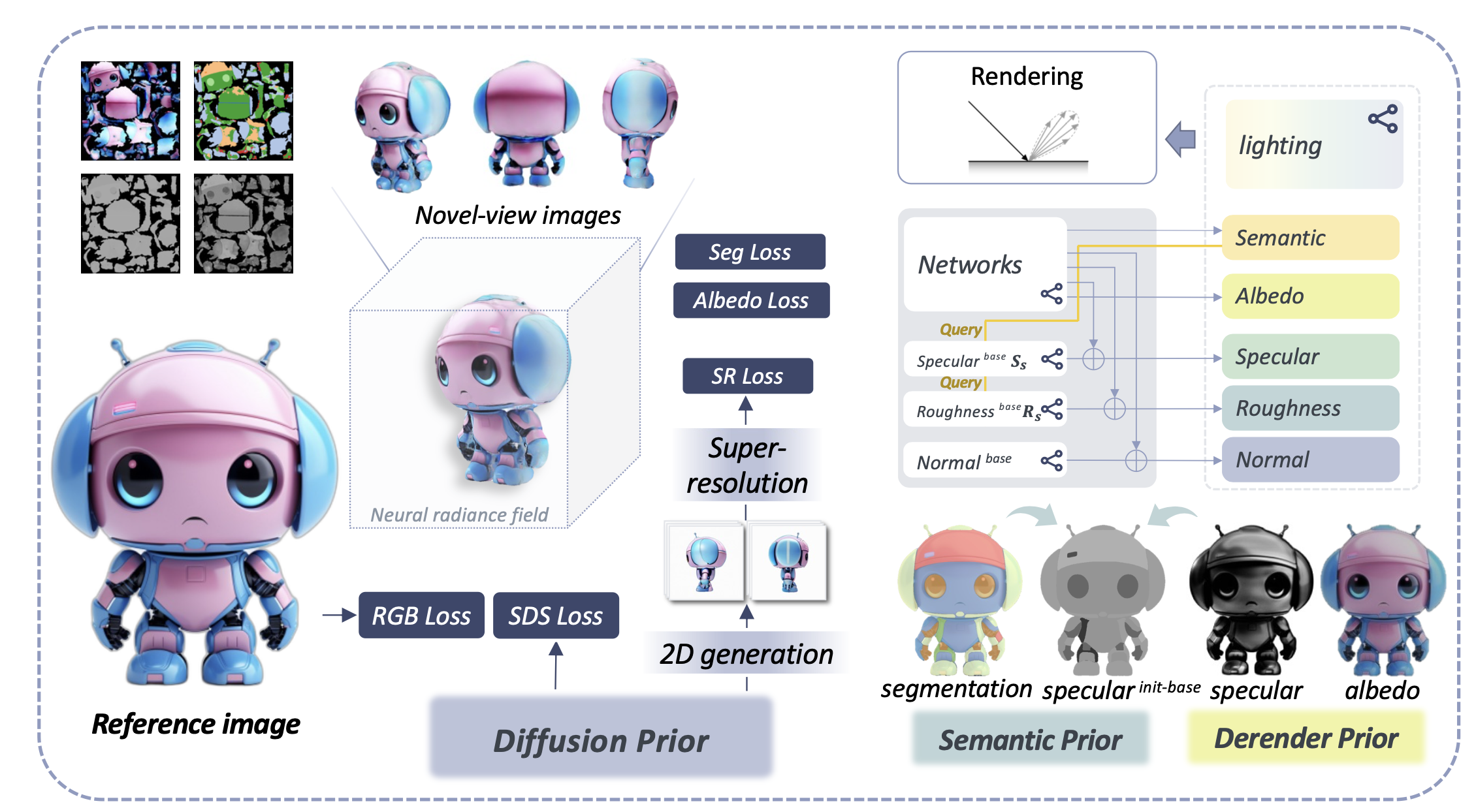}
    \caption{Albedo Aware 3D Content Generation by HyperDreamer \cite{wu_hyperdreamer_2023}.}
    \label{fig:hyperdreamer-albedo}
\end{figure}

Meanwhile, HyperDreamer: Hyper-Realistic 3D Content Generation and Editing from a Single Image \cite{wu_hyperdreamer_2023} exemplifies the integration of 2D backbones into 3D workflows, as depicted in Figure \ref{fig:hyperdreamer-albedo}, with a focus on regulating albedo \footnote{Intrinsic color of a surface, representing how much light it reflects diffusely. It excludes shading, shadows, or lighting effects and is commonly used in 3D graphics and rendering to define surface appearance.} for photorealistic rendering. Finally, Automated Video Editing Based on Learned Styles Using LSTM-GAN \cite{huang_automated_2022} explores a novel approach to video editing using LSTM-GANs.

\subsection{Effective Architectures for Generating Congruent 3D Worlds}

For the works identified in the literature review, table \ref{tab:model_references} maps the most adopted architectural paradigms for generating three-dimensional outputs, including meshes, point clouds, and voxel shapes.

\begin{table}[H]
\centering
\renewcommand{\arraystretch}{1.2} 
\setlength{\tabcolsep}{5pt} 
\begin{tabular}{|p{3cm}|p{5cm}|} 
\hline
Model & References \\ 
\hline
Diffusion Model & \cite{liu_3dall-e_2023}, \cite{zhao_vaide_2024}, \cite{wang_roomdreaming_2024}, \cite{he_creating_2024}, \cite{kim_neuralfield-ldm_2023}, \cite{dong_coin3d_2024}, \cite{hu_cns-edit_2024}, \cite{li_icontrol3d_2024}, \cite{chen_memovis_2024}, \cite{zhang_protodreamer_2024}, \cite{rajaram_blendscape_2024}, \cite{huang_blue_2024}, \cite{alaluf_neural_2023}, \cite{chen_find_2024}, \cite{richardson_conceptlab_2024}, \cite{li_sat2scene_2024}, \cite{zuo_dg3d_2023}, \cite{huang_plantography_2024}, \cite{yiyang_context-aware_2024}, \cite{han_scene_2024}, \cite{petrov_gem3d_2024}, \cite{mohammad_khalid_clip-mesh_2022}, \cite{dang_worldsmith_2023}, \cite{si_generating_2024}, \cite{shen_neural_2024}, \cite{wu_blockfusion_2024}, \cite{cai_l-magic_2024}, \cite{kant_invs_2023}, \cite{zhang_clay_2024}, \cite{zhou_unsupervised_2022}, \cite{ryu_360_2023}, \cite{perla_easi-tex_2024}, \cite{zhang_3dshape2vecset_2023}, \cite{wu_hyperdreamer_2023}, \cite{xiao_text-free_2024}, \cite{guerrero-viu_texsliders_2024}, \cite{sella_spicee_2024}, \cite{zhang_text--3d_2023}, \cite{li_imagebind3d_2024}, \cite{zhong_dreamlcm_2024}, \cite{weng_dream_2024}, \cite{huang_placiddreamer_2024}, \cite{yin_text2vrscene_2024}, \cite{chen_controlstyle_2023}, \cite{vecchio_controlmat_2024}, \cite{hu_diffusion_2024}, \cite{zhang_mapa_2024}, \cite{rios_large_2023}, \cite{zhang_texpainter_2024},
\cite{wang_themestation_2024} \\ 
VAE & \cite{gao_scenehgn_2023}, \cite{wu_dreamup3d_2024}, \cite{zhang_clay_2024}, \cite{nakada_study_2023}, \cite{li_cad_2024}, \cite{huang_vdam_2022}, \cite{lin_cascade_2022}, \cite{faivishevsky_automated_2021} \\ 
GAN & \cite{jabbar_survey_2021}, \cite{zuo_dg3d_2023}, \cite{son_singraf_2023}, \cite{awiszus_world-gan_2021}, \cite{rai_assessing_2024}, \cite{chen_shaddr_2023}, \cite{wu_learning_2022}, \cite{li_imagebind3d_2024}, \cite{upadhyay_floorgan_2023}, \cite{faivishevsky_automated_2021} \\ 
Gaussian Splatting & \cite{sun_f-3dgs_2024} \\
Flow-based & \cite{kimura_chartpointflow_2021} \\
\hline
\end{tabular}
\caption{Works Classified According to Different Architectures}
\label{tab:model_references}
\end{table}

Diffusion models emerge as a robust backbone for 3D object generation, showcasing their ability to model complex data distributions and generate high-quality results. GANs and their variants, such as WGAN \cite{merino_interactive_2023}, SuperGAN \cite{maesumi_explorable_2023}, and AttentionGAN \cite{harris_multi-stage_2021}, were frequently employed because of their ability to produce realistic 3D content through adversarial training, at the cost of risking mode collapse. VAEs proved useful in learning compact latent representations \cite{nakada_study_2023}, facilitating efficient data encoding and reconstruction. 

Other innovative approaches also gained attention. Gaussian splatting demonstrated exceptional efficiency in reconstructing objects \cite{sun_f-3dgs_2024}, overcoming some limitations of traditional neural rendering techniques. Flow-based methods \cite{kimura_chartpointflow_2021} offered improvements in preserving the topology of rendered objects, ensuring structural consistency in outputs. Additionally, Monte Carlo Markov Chains, combined with Metropolis-Hastings kernels, achieved impressive results in probabilistically modeling object distributions within scenes \cite{gothoskar_3dp3_2021}. These architectures collectively provide a comprehensive foundation for addressing the challenges of 3D generation, and their implications will be further discussed in the context of this thesis.

\begin{figure}[H]
    \centering
    \includegraphics[width=1\linewidth]{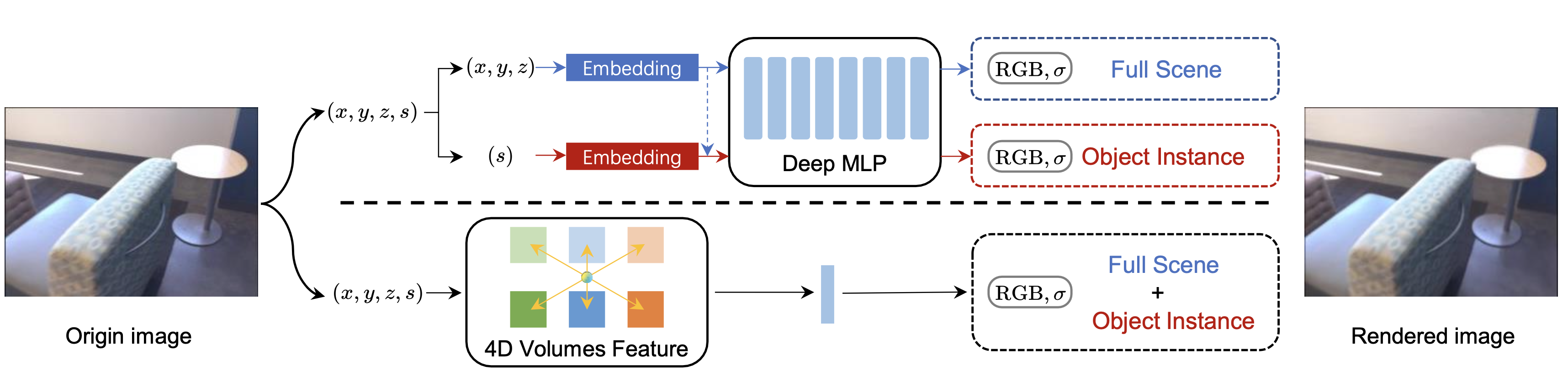}
    \caption{NeRF-IS \cite{sha_nerf-is_2024}: Encoding Semantic Information in Diffusion Process.}
    \label{fig:nerfis-encoding}
\end{figure}

Architectures integrating advanced radiance fields and scene conditioning further enrich this space. NeRF-IS \cite{sha_nerf-is_2024}, which architecture is represented in Figure \ref{fig:nerfis-encoding}, encodes semantic information directly into neural radiance fields, making diffusion spaces more interpretable and structured.

SceneHGN: Hierarchical Graph Networks for 3D Indoor Scene Generation With Fine-Grained Geometry \cite{gao_scenehgn_2023} utilizes multi-stage training strategies to produce stable and detailed indoor scenes, pushing the boundaries of architectural and object-level precision. Scene Diffusion: Text-driven Scene Image Synthesis Conditioning on a Single 3D Model
\cite{han_scene_2024} bridges textual prompts and scene synthesis by conditioning diffusion processes on single 3D models, enhancing contextual relevance in generated environments.

Several methods prioritize controllability and thematic flexibility in 3D generation. For instance, CLAY: A Controllable Large-scale Generative Model for Creating High-quality 3D Assets, integrates diffusion transformers to separate and refine mesh and terrain assets, enhancing asset specificity \cite{zhang_clay_2024}. Similarly, Coin3D \cite{dong_coin3d_2024} uses proxy-guided conditioning to generate controllable and interactive 3D assets, opening new research streams for user-driven terrain and structural generation. ThemeStation \cite{wang_themestation_2024} employs dual-score distillation to create theme-aware 3D assets, ensuring coherence and creativity with minimal exemplars. On a broader scale, World-GAN \cite{awiszus_world-gan_2021} introduces a method to modify styles post-training via latent space interpretation, offering new possibilities for creative manipulation of large-scale 3D terrains, such as Minecraft-like worlds.

With respect to output types, table \ref{tab:output_references} maps the reviews work landscape, where there is a predominance of geometric mesh output. 
\begin{table}[ht!]
\centering
\renewcommand{\arraystretch}{1.2} 
\begin{tabular}{|p{3cm}|p{5cm}|} %
\hline
Output & References \\ 
\hline
Image & \cite{liu_3dall-e_2023}, \cite{kikuchi_constrained_2021}, \cite{chen_memovis_2024}, \cite{harris_multi-stage_2021}, \cite{rajaram_blendscape_2024}, \cite{jabbar_survey_2021}, \cite{rosenberg_drawtalking_2024}, \cite{huang_blue_2024}, \cite{alaluf_neural_2023}, \cite{chung_artinter_2023}, \cite{cao_autoencoder-based_2024}, \cite{chen_find_2024}, \cite{richardson_conceptlab_2024}, \cite{han_scene_2024}, \cite{dang_worldsmith_2023}, \cite{si_generating_2024}, \cite{cai_l-magic_2024}, \cite{kant_invs_2023}, \cite{chen_controlstyle_2023},
\cite{tang_realfill_2024}, \cite{xia_survey_2023}, \cite{song_image_2023} \\ 
Mesh & \cite{sha_nerf-is_2024}, \cite{zhao_vaide_2024}, \cite{hertz_spaghetti_2022}, \cite{hu_generative_2024}, \cite{li_sp-gan_2021}, \cite{he_creating_2024}, \cite{de_la_torre_llmr_2024}, \cite{kim_neuralfield-ldm_2023}, \cite{dong_coin3d_2024}, \cite{hu_cns-edit_2024}, \cite{maesumi_explorable_2023}, \cite{li_icontrol3d_2024}, \cite{wang_proteusnerf_2024}, \cite{yang_geolatent_2023}, \cite{li_sat2scene_2024}, \cite{jiang_synthesizing_2021}, \cite{zuo_dg3d_2023}, \cite{mo_sparse_2024}, \cite{yiyang_context-aware_2024}, \cite{son_singraf_2023}, \cite{merino_interactive_2023}, \cite{awiszus_world-gan_2021}, \cite{petrov_gem3d_2024}, \cite{mohammad_khalid_clip-mesh_2022}, \cite{zhang_vrcopilot_2024}, \cite{shen_neural_2024}, \cite{wu_blockfusion_2024}, \cite{singh_worldgen_2023}, \cite{jones_shapecoder_2023}, \cite{rai_assessing_2024}, \cite{chen_shaddr_2023}, \cite{chen_lart_2023}, \cite{wu_dreamup3d_2024}, \cite{zhang_clay_2024}, \cite{ma_multicad_2023}, \cite{zhou_unsupervised_2022}, \cite{ryu_360_2023}, \cite{cheng_cross-modal_2022}, \cite{yan_online_2021}, \cite{wang_elevating_2023}, \cite{perla_easi-tex_2024}, \cite{sun_f-3dgs_2024}, \cite{zhang_3dshape2vecset_2023}, \cite{wu_hyperdreamer_2023}, \cite{zhang_end--end_2022}, \cite{wu_learning_2022}, \cite{xiao_text-free_2024}, \cite{wang_embracing_2021}, \cite{faruqi_style2fab_2023}, \cite{zhou_edit3d_2024}, \cite{guerrero-viu_texsliders_2024}, \cite{sella_spicee_2024}, \cite{zhang_text--3d_2023}, \cite{ye_relscene_2024}, \cite{li_imagebind3d_2024}, \cite{zhong_dreamlcm_2024}, \cite{weng_dream_2024}, \cite{huang_placiddreamer_2024}, \cite{yin_text2vrscene_2024}, \cite{polys_prompt_2024}, \cite{zheng_sketch3d_2024}, \cite{numan_spaceblender_2024}, \cite{vecchio_controlmat_2024}, \cite{hu_diffusion_2024}, \cite{bonic_broomrocket_2024}, \cite{guerrero_matformer_2022}, \cite{zhang_mapa_2024}, \cite{zhou_photomat_2023}, \cite{rios_large_2023}, \cite{zhang_texpainter_2024}, \cite{wang_themestation_2024}, \cite{agrawal_image_2024} \\  
Voxel & \cite{nakada_study_2023} \\ 
Video & \cite{huang_automated_2022} \\ 
Point Cloud & \cite{kimura_chartpointflow_2021} \\ 
Graph & \cite{upadhyay_floorgan_2023} \\ 
Text & \cite{lee_knowledge_2023}, \cite{huang_vdam_2022} \\ 
\hline
\end{tabular}
\caption{Works classified by the output of their architecture}
\label{tab:output_references}
\end{table}

\subsection{Techniques to Coherently Blend Elements Aesthetics}

\begin{figure}[H]
    \centering
    \includegraphics[width=1\linewidth]{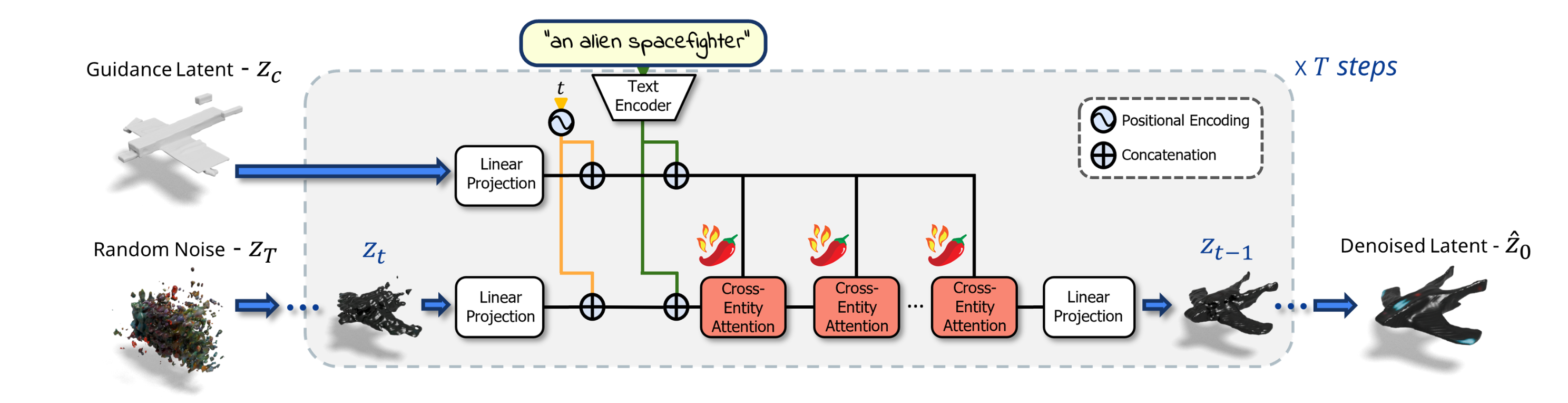}
    \caption{Usage of Cross-Attention in Spice-E \cite{sella_spicee_2024} Architecture.}
    \label{fig:spice-attention}
\end{figure}

New style transfer approaches enable the efficient production of assets in diverse styles. Diffusion Transformers \cite{zhang_clay_2024} address the style transfer process by iteratively refining outputs and disentangling style and content features over adaptive latent sizes. Cross-attention mechanisms ensure that stylistic transformations are applied contextually and uniformly \cite{zhou_edit3d_2024}, maintaining coherence and realism across multiple perspectives. Latent space representations \cite{yang_geolatent_2023} compress high-dimensional data into manageable forms, allowing for precise manipulation of stylistic attributes while preserving the geometry and texture of 3D models.  Figure \ref{fig:spice-attention} illustrates Spice-E's \cite{sella_spicee_2024} pipeline, which employs cross-attention to shift the output of the generative model and align it with the textual prompt that encodes the desired style.

\begin{figure}[H]
    \centering
    \includegraphics[width=0.7\linewidth]{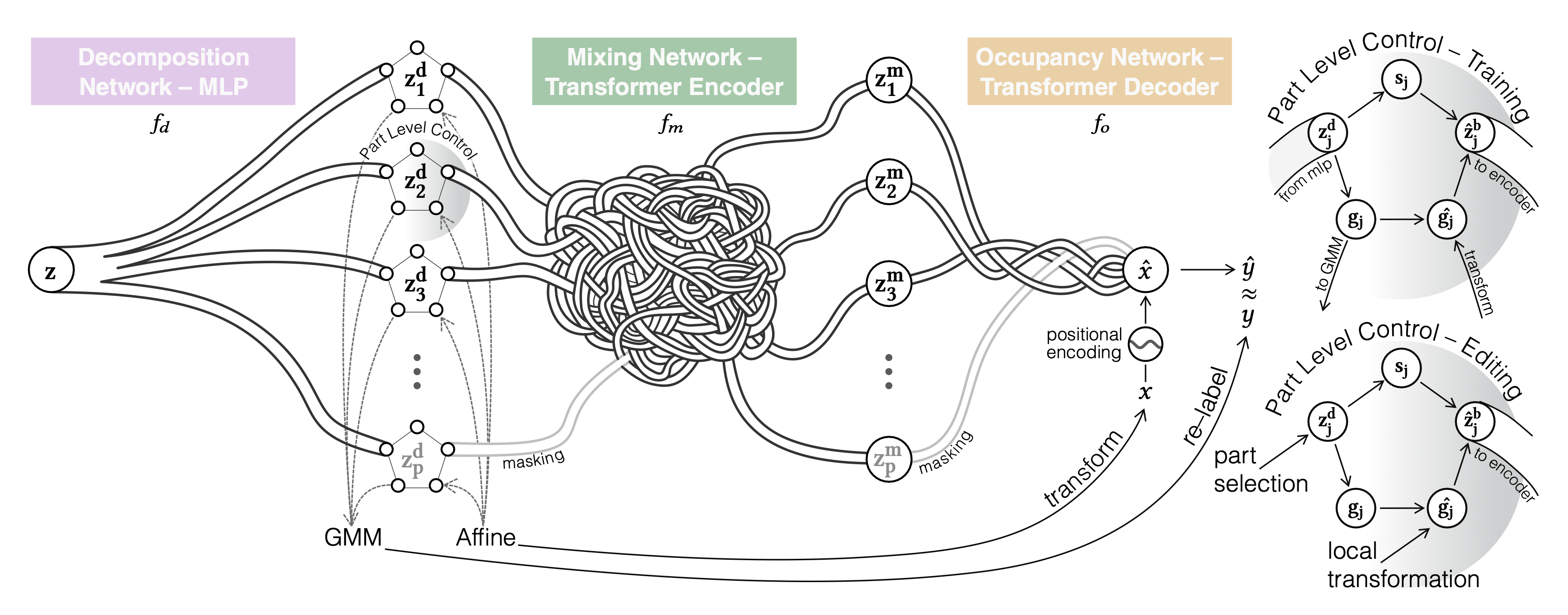}
    \caption{SPAGHETTI \cite{hertz_spaghetti_2022} architecture enhances Control over Shape Parts.}
    \label{fig:spaghetti-decoupling}
\end{figure}

Diffusion models incorporating innovative approaches to noise and optimization mechanisms have also emerged as pivotal tools. Blue noise for diffusion models \cite{huang_blue_2024} explores novel noise signals, such as blue Noise, which enhances detail retention in the diffusion process, particularly in image generation. Complementary to this, CNS-Edit \cite{hu_cns-edit_2024} introduces Coupled Neural Shape optimization, enabling precise 3D shape editing. Techniques like SPAGHETTI \cite{hertz_spaghetti_2022}, highlighted in Figure \ref{fig:spaghetti-decoupling}, which combines implicit shape editing with part-aware generation, employ Gaussian Mixture Models (GMMs) as intermediate representations, paving the way for modular and controllable shape manipulation. Self-supervised learning on 3D Point Clouds by Learning Discrete Generative Models \cite{eckart_self-supervised_2021} uses GMMs as autoencoder bottlenecks to learn more structured and self-supervised point cloud representations, bridging the gap between latent encodings and 3D reconstruction.

\subsection{Limitations in lifelike 3D scene generation}

The limitations of current AI models in the generation of lifelike 3D scenes stem from several critical factors, one of them being the immense processing power needed to run most of these applications. Additionally, achieving high resolution and fine details remains challenging, with many models struggling to produce outputs that meet the visual quality expected for lifelike environments \cite{rai_assessing_2024}. These shortcomings are exacerbated by the lack of diverse and high-quality datasets or larger-scale real-world data \cite{kim_neuralfield-ldm_2023} that fit needed generative tasks, like 3D editing, which limits the ability of models to generalize and create varied and realistic scenes.

Another prominent issue is the difficulty in handling the robustness and reliability of generated outputs. Current models often encounter challenges in maintaining consistent quality, particularly when working with limited or unevenly distributed training data \cite{wang_roomdreaming_2024}. Generalization issues are also common, where models fail to adapt to unseen scenarios or complex environments. Furthermore, real-time applications remain largely out of reach due to the computational complexity of most approaches \cite{wu_hyperdreamer_2023}. Addressing these limitations will require advancements in model architectures \cite{yin_text2vrscene_2024}, optimization techniques, and access to better datasets to push the boundaries of 3D scene generation.

For unbounded environment generation, the challenges of scalability and generalization become critical in comparing between approaches. GANs, while capable of producing sharp outputs, struggle with maintaining diversity in large or complex scenes, making them less ideal for expansive environments \cite{son_singraf_2023}. VAEs, though computationally efficient, are limited in detail and scalability, which constrains their ability to model unbounded environments effectively, performing better in lower-dimensional spaces \cite{huang_vdam_2022}. Diffusion models, with their ability to capture fine details and realism, are better suited for generating large-scale, unbounded scenes, but their high computational requirements and slower training times remain significant hurdles \cite{wang_themestation_2024}.

\subsection{Balancing the influence of multiple types of input}

\begin{figure}
    \centering
    \includegraphics[width=1\linewidth]{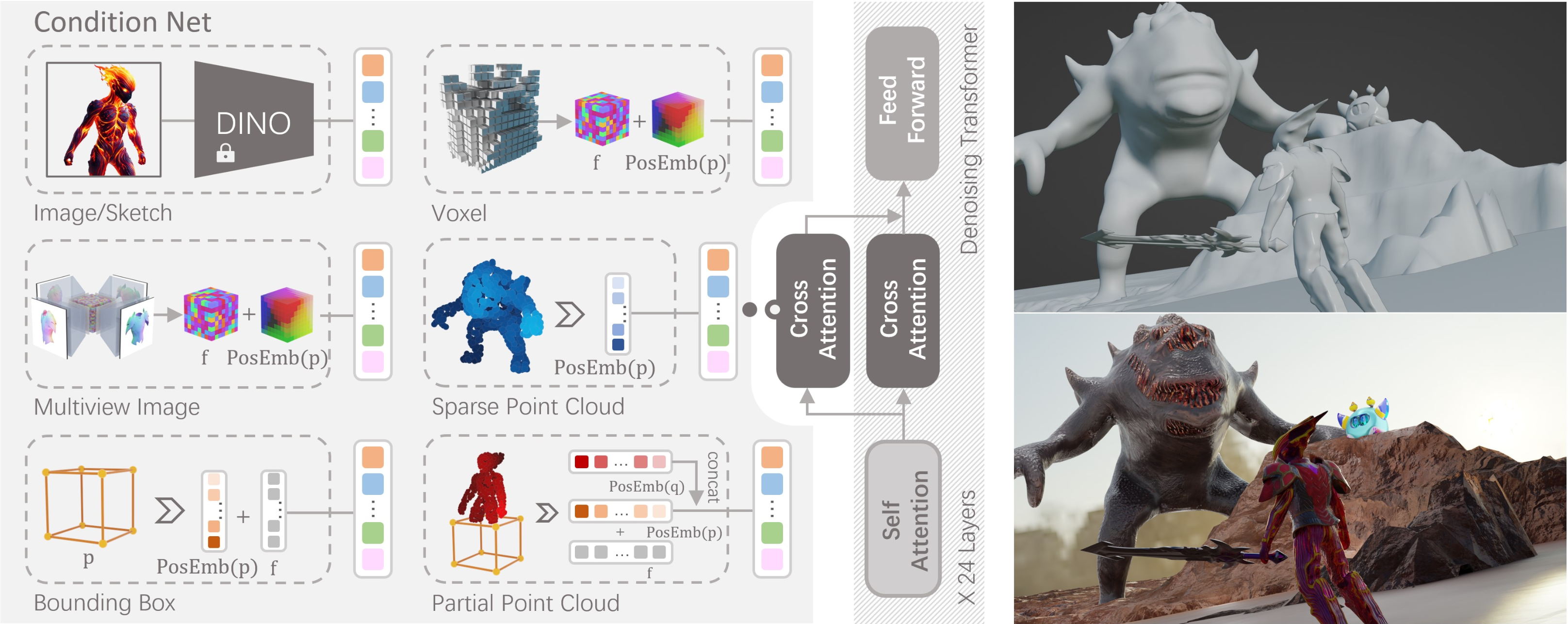}
    \caption{3D Scene Generation Pipeline from CLAY \cite{zhang_clay_2024} with different Input Modalities.}
    \label{fig:clay-generation}
\end{figure}

Gen AI balances contextual 3D artifacts and textual prompts through advanced multi-modal integration techniques that align and fuse information from different modalities. Cross-attention and multi-modal attention are central to this process, enabling the model to relate textual inputs to 3D features by attending to the most relevant aspects of each. Techniques like multimodal embeddings \cite{liu_3dall-e_2023} and spatial-semantic joint space representations \cite{sha_nerf-is_2024} unify textual and visual data into a shared space, ensuring coherence between prompts and 3D outputs. Meanwhile, proxy-guided conditioning \cite{dong_coin3d_2024} and layout-guided diffusion \cite{wu_blockfusion_2024} allows the model to anchor textual inputs to specific 3D spatial structures, creating environments that adhere to both semantic and spatial requirements, such as CLAY \cite{zhang_clay_2024}, which is illustrated in Figure \ref{fig:clay-generation}. Techniques like edge-map conditioning \cite{perla_easi-tex_2024} and BMIC with depth maps \cite{zhang_end--end_2022} further reinforce geometric accuracy, ensuring that generated artifacts remain structurally consistent with textual prompts.

\begin{table}[H]
\centering
\renewcommand{\arraystretch}{1.3} 
\small 
\begin{tabularx}{\textwidth}{|p{3cm}|p{5cm}|}
\hline
Approach & References \\ 
\hline
Multi-Modal Embeddings & \cite{liu_3dall-e_2023}, \cite{mohammad_khalid_clip-mesh_2022} \\
Cross-Attention & \cite{kim_neuralfield-ldm_2023}, \cite{harris_multi-stage_2021}, \cite{petrov_gem3d_2024}, \cite{ryu_360_2023}, \cite{zhang_3dshape2vecset_2023}, \cite{zhou_edit3d_2024}, \cite{li_imagebind3d_2024} \\
Textual Bypass & \cite{alaluf_neural_2023} \\ 
Cascading Contrastive Strategy & \cite{li_cad_2024} \\
Optimal Transport & \cite{ma_multicad_2023} \\ 
Joint Space Representation & \cite{sha_nerf-is_2024} \\ 
\hline
\end{tabularx}
\caption{Works classified by the approach used to blend multi-modal inputs}
\label{tab:multi_modal}
\end{table}

To refine the alignment and improve generation quality, techniques such as cascading contrastive strategies \cite{li_cad_2024} and latent space alignment ensure that the relationship between modalities is preserved at multiple levels of detail. PixelCNN \cite{nakada_study_2023} latent sampling and multi-scale image encoders help process inputs at different granularities, capturing both global context and fine details. Additionally, methods like optimal transport \cite{ma_multicad_2023} align the statistical distributions of textual and visual data, ensuring that features from both modalities influence the output harmoniously, effectively bridging the gap between textual and visual domains.

Other techniques also play crucial roles in enhancing the balance between textual prompts and 3D artifacts. Textual bypass \cite{alaluf_neural_2023} mechanisms provide a direct influence of text on the generation process, ensuring that prompts are explicitly reflected in the output without being diluted by intermediate layers. Conversion of neural material features into analytical reflectance maps \cite{zhou_photomat_2023} contributes to the physical realism of 3D scenes by accurately modeling how surfaces interact with light. For capturing large-scale scenes or diverse input types, a combination of different visual inputs and 2D exploration space approaches allow the model to process multiple layers of visual information, while cross-entity attention focuses on understanding relationships between objects within a scene, enhancing spatial coherence. Lastly, tools like PCA simplify complex data inputs for efficient processing, and multi-scale image encoders ensure that fine details and broader context are simultaneously integrated into the generative process, reinforcing both textual and visual fidelity in the final outputs. These complementary techniques collectively expand the model's ability to produce well-aligned, high-quality 3D environments.

\subsection{Impact in the diversity of datasets}

The quality and diversity of training data significantly influence the output quality of generated 3D environments. High-quality datasets with diverse modalities (e.g., RGB, TEXT, MESH) allow models to generalize better and produce rich, multifaceted outputs \cite{cheng_cross-modal_2022}. For instance, datasets like ShapeNet (leveraged in 15 papers) and Objaverse (used in 7 papers) contribute to realistic 3D reconstructions and novel view synthesis. However, limited diversity in data or noisy datasets can result in outputs that lack adaptability, detail, or realism in novel scenarios \cite{wang_roomdreaming_2024}. Metrics like IoU, Chamfer Distance, and FID benefit from richer datasets, reflecting enhanced generalization and visual fidelity. To mitigate the challenges of training from scratch, many models incorporate pre-trained embeddings, leveraging pre-existing knowledge to improve efficiency and accuracy.

\begin{table*}
\centering
\renewcommand{\arraystretch}{1.3} 
\small 
\begin{tabularx}{\textwidth}{X X}
\hline
Input & References \\ 
\hline
Text & \cite{liu_3dall-e_2023}, \cite{he_creating_2024}, \cite{kikuchi_constrained_2021}, \cite{li_icontrol3d_2024}, \cite{chen_memovis_2024}, \cite{zhang_protodreamer_2024}, \cite{harris_multi-stage_2021}, \cite{jabbar_survey_2021}, \cite{chung_artinter_2023}, \cite{chen_find_2024}, \cite{richardson_conceptlab_2024}, \cite{huang_plantography_2024}, \cite{mohammad_khalid_clip-mesh_2022}, \cite{dang_worldsmith_2023}, \cite{si_generating_2024}, \cite{shen_neural_2024}, \cite{cai_l-magic_2024}, \cite{rai_assessing_2024}, \cite{zhang_clay_2024}, \cite{cheng_cross-modal_2022}, \cite{wang_elevating_2023}, \cite{zhang_3dshape2vecset_2023}, \cite{faruqi_style2fab_2023}, \cite{guerrero-viu_texsliders_2024}, \cite{sella_spicee_2024}, \cite{zhang_text--3d_2023}, \cite{ye_relscene_2024}, \cite{li_cad_2024}, \cite{li_imagebind3d_2024}, \cite{zhong_dreamlcm_2024}, \cite{weng_dream_2024}, \cite{huang_placiddreamer_2024}, \cite{yin_text2vrscene_2024}, \cite{polys_prompt_2024}, \cite{chen_controlstyle_2023}, \cite{vecchio_controlmat_2024}, \cite{bonic_broomrocket_2024}, \cite{rios_large_2023}, \cite{upadhyay_floorgan_2023}, \cite{zhang_texpainter_2024}, \cite{son_genquery_2024}, \cite{chung_promptpaint_2023} \\ 
Image & \cite{liu_3dall-e_2023}, \cite{rajaram_blendscape_2024}, \cite{ryu_360_2023}, \cite{zhang_3dshape2vecset_2023},
\cite{zhao_vaide_2024}, \cite{wang_roomdreaming_2024}, \cite{he_creating_2024}, \cite{kim_neuralfield-ldm_2023}, \cite{wang_proteusnerf_2024}, \cite{huang_blue_2024}, \cite{li_sat2scene_2024}, \cite{zuo_dg3d_2023}, \cite{gottsacker_rlty2rlty_2024}, \cite{han_scene_2024}, \cite{son_singraf_2023}, \cite{shen_neural_2024}, \cite{kant_invs_2023}, \cite{zhang_clay_2024}, \cite{zhou_unsupervised_2022}, \cite{cheng_cross-modal_2022}, \cite{wang_elevating_2023}, \cite{de_sousa_ribeiro_object-centric_2024}, \cite{xiao_text-free_2024}, \cite{wang_embracing_2021}, \cite{zhou_edit3d_2024}, \cite{numan_spaceblender_2024}, \cite{hu_diffusion_2024}, \cite{lin_cascade_2022}, \cite{tang_realfill_2024}, \cite{son_genquery_2024}\\ 
Text and Image & \cite{liu_3dall-e_2023}, \cite{rajaram_blendscape_2024}, \cite{jabbar_survey_2021}, \cite{rosenberg_drawtalking_2024}, \cite{alaluf_neural_2023}, \cite{chung_artinter_2023},
\cite{li_imagebind3d_2024}, \cite{zheng_sketch3d_2024},  \cite{chen_controlstyle_2023}, \cite{vecchio_controlmat_2024}\\ 
Image and Depth & \cite{kim_neuralfield-ldm_2023}, \cite{gothoskar_3dp3_2021}, \cite{wu_dreamup3d_2024} \\ 
Mesh & \cite{dong_coin3d_2024}, \cite{hu_cns-edit_2024}, \cite{yang_geolatent_2023}, \cite{wu_blockfusion_2024}, \cite{singh_worldgen_2023}, \cite{perla_easi-tex_2024}, \cite{zhang_3dshape2vecset_2023}, \cite{wu_hyperdreamer_2023}, \cite{zhang_end--end_2022}, \cite{wu_learning_2022}, \cite{lee_knowledge_2023}, \cite{zhang_texpainter_2024}, \cite{wang_themestation_2024} \\ 
Voxel & \cite{awiszus_world-gan_2021}, \cite{chen_shaddr_2023}, \cite{nakada_study_2023} \\ 
Video & \cite{huang_automated_2022}, \cite{faivishevsky_automated_2021} \\ 
Point Cloud & \cite{mo_sparse_2024}, \cite{petrov_gem3d_2024}, \cite{ma_multicad_2023}, \cite{kimura_chartpointflow_2021}, \cite{yan_online_2021}, \cite{eckart_self-supervised_2021}, \cite{sun_f-3dgs_2024} \\ 
Point Cloud and Text & \cite{yiyang_context-aware_2024} \\  

Graph & \cite{upadhyay_floorgan_2023} \\ 
Height Map & \cite{hu_generative_2024} \\ 
Text and Mesh & \cite{zhang_mapa_2024} \\ 

\hline
\end{tabularx}
\caption{Most used Inputs across all references}
\label{tab:inputs_used}
\end{table*}

The widespread use of pre-trained embeddings highlights their practical advantages in tasks involving 3D generation. These embeddings, derived from large, diverse datasets, enable faster development and better performance by providing a strong foundation for downstream tasks. Pre-trained models reduce computational costs, capture rich representations, and generalize effectively across domains. By utilizing these pre-trained resources, researchers can focus on fine-tuning for specific applications, such as AR/VR simulations, gaming, and interactive environments, without the need for extensive training from scratch. This approach accelerates innovation and enhances scalability, ensuring high-quality outputs while addressing computational and resource limitations. 

Discrepancies are still evident in the most commonly used modalities and datasets (table \ref{tab:inputs_used}). While certain modalities like images, text, and meshes were frequently utilized, their adoption varied significantly across studies depending on the application focus. This variation highlights a lack of standardization in the selection of modalities and datasets, which can lead to differences in performance, generalization, and applicability between models.

Regarding ways of representing shapes, techniques like 3DShape2VecSet \cite{zhang_3dshape2vecset_2023}, leverage neural fields and generative diffusion models to facilitate high-quality reconstruction and creation of detailed assets. Meanwhile, A Study on Voxel Shape Generation and Reconstruction with VQ-VAE-2 \cite{nakada_study_2023} reveals a novel widening approach to enhance detail retention in voxel-based reconstructions, advancing the fidelity of 3D generative processes.

\begin{figure}[H]
    \centering
    \includegraphics[width=1\linewidth]{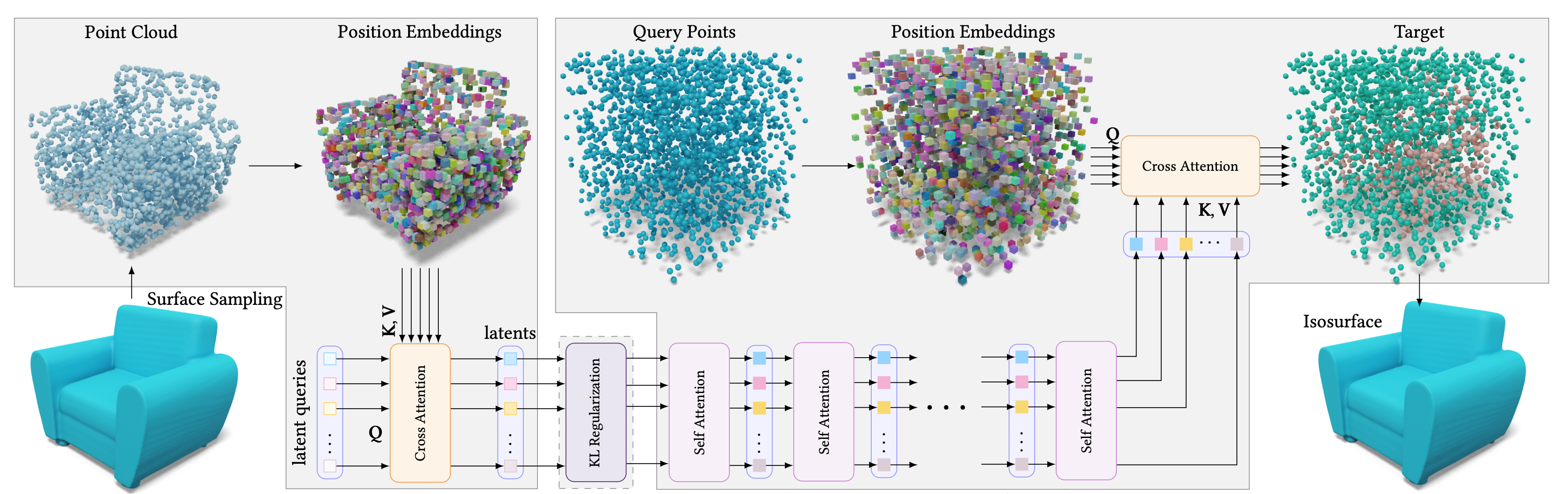}
    \caption{Shaping Autoencoding Pipeline in 3DShape2VecSet \cite{zhang_3dshape2vecset_2023}.}
    \label{fig:autoencode-3d2vec}
\end{figure}

\subsection{Criteria for assessing generated environments}

Most of the metrics mentioned in the literature can be broadly categorized into geometric or perceptual measures:

\subsubsection{Geometrical Quality Metrics}
These metrics directly evaluate the structure and spatial accuracy of the generated 3D scene (often by comparing point clouds, meshes, or volumetric representations):
\begin{itemize}
    \item Intersection over Union (IoU): Measures the overlap between generated and ground truth volumes.
    \item Chamfer Distance: Quantifies the average distance between points in the generated and reference point clouds.
    \item F-Score: Balances precision and recall in matching points or surface elements.
    \item Frechet Point Distance: A variant of the Frechet distance applied to point clouds.
    \item Reconstruction Error: Directly measures the difference between generated 3D geometry and the ground truth.
    \item 1-NNA (1-Nearest Neighbor Accuracy): Evaluates distribution similarity (often to detect overfitting or mode collapse in the geometric domain).
    \item EMD (Earth Mover’s Distance): Assesses the minimal “cost” to transform one point distribution into another.
\end{itemize}

\subsubsection{Perceptual Quality Metrics}

These metrics assess the quality of rendered images or the perceptual similarity of the generated scenes compared to real examples. They often involve comparisons of learned feature distributions or human-perceived image quality:
\begin{itemize}
    \item Frechet Inception Distance (FID): Compares feature distributions (usually from a pretrained Inception network) of generated and real images.
    \item Kernel Inception Distance: Similar to FID, but uses kernel methods to compare distributions.
    \item PSNR (Peak Signal-to-Noise Ratio): Measures pixel-level fidelity of rendered images.
    \item SSIM (Structural Similarity Index): Evaluates the structural similarity between images.
    \item LPIPS (Learned Perceptual Image Patch Similarity): Uses deep features to capture perceptual differences.
    \item CLIP Similarity: Leverages CLIP embeddings to evaluate how well the generated scene matches a given textual or visual prompt.
    \item MMD (Maximum Mean Discrepancy): Often used to compare the statistical distribution of generated features with real data, which can apply in a perceptual context.
\end{itemize}

Table \ref{tab:metric_references} highlights a range of metrics used to evaluate the authenticity of independently generated environments, highlighting their application in Gen AI for the creation of 3D assets. 

\begin{table}[ht!]
\centering
\renewcommand{\arraystretch}{1.2} 
\begin{tabular}{|p{3cm}|p{5cm}|} 

\hline
Metric & References \\ 
\hline
IoU & \cite{kikuchi_constrained_2021}, \cite{chen_shaddr_2023}, \cite{wu_dreamup3d_2024}, \cite{zhang_clay_2024}, \cite{eckart_self-supervised_2021}, \cite{zhang_3dshape2vecset_2023}, \cite{wu_learning_2022}, \cite{nakada_study_2023}, \cite{zhou_edit3d_2024} \\ 
Chamfer Distance & \cite{zuo_dg3d_2023}, \cite{petrov_gem3d_2024}, \cite{wu_blockfusion_2024}, \cite{zhang_clay_2024}, \cite{ma_multicad_2023}, \cite{cheng_cross-modal_2022}, \cite{zhang_3dshape2vecset_2023}, \cite{rios_large_2023}, \cite{xia_survey_2023} \\ 
F-Score & \cite{chen_shaddr_2023}, \cite{zhang_clay_2024}, \cite{zhang_3dshape2vecset_2023}, \cite{wu_learning_2022}, \cite{wang_controllable_2024} \\ 
FID & \cite{zhao_vaide_2024}, \cite{kim_neuralfield-ldm_2023}, \cite{hu_cns-edit_2024}, \cite{kikuchi_constrained_2021}, \cite{gao_scenehgn_2023}, \cite{maesumi_explorable_2023}, \cite{jabbar_survey_2021}, \cite{huang_blue_2024}, \cite{cao_autoencoder-based_2024}, \cite{yang_geolatent_2023}, \cite{li_sat2scene_2024}, \cite{zuo_dg3d_2023}, \cite{petrov_gem3d_2024}, \cite{chen_shaddr_2023}, \cite{zhang_clay_2024}, \cite{zhou_unsupervised_2022}, \cite{cheng_cross-modal_2022}, \cite{zhang_3dshape2vecset_2023}, \cite{wu_learning_2022}, \cite{ye_relscene_2024}, \cite{li_imagebind3d_2024}, \cite{hu_diffusion_2024}, \cite{guerrero_matformer_2022}, \cite{zhang_mapa_2024}, \cite{zhou_photomat_2023}, \cite{upadhyay_floorgan_2023}, \cite{zhang_texpainter_2024}, \cite{lin_cascade_2022}, \cite{xia_survey_2023}, \cite{kaswan_generative_2023}, \cite{yang_diffusion_2023}, \cite{wang_generative_2021} \\ 
KID & \cite{hu_cns-edit_2024}, \cite{li_sat2scene_2024}, \cite{son_singraf_2023}, \cite{petrov_gem3d_2024}, \cite{zhang_clay_2024}, \cite{zhang_3dshape2vecset_2023}, \cite{ye_relscene_2024}, \cite{zhang_mapa_2024}, \cite{xia_survey_2023}, \cite{yang_diffusion_2023} \\ 
FPD & \cite{li_sp-gan_2021}, \cite{zhang_3dshape2vecset_2023}, \cite{xiao_text-free_2024}, \cite{li_imagebind3d_2024} \\ 
PSNR & \cite{sha_nerf-is_2024}, \cite{harris_multi-stage_2021}, \cite{huang_blue_2024}, \cite{cao_autoencoder-based_2024}, \cite{li_sat2scene_2024}, \cite{wang_rip-nerf_2023}, \cite{kant_invs_2023}, \cite{zhou_unsupervised_2022}, \cite{ryu_360_2023}, \cite{cheng_cross-modal_2022}, \cite{sun_f-3dgs_2024}, \cite{tang_realfill_2024}, \cite{xia_survey_2023} \\ 
SSIM & \cite{sha_nerf-is_2024}, \cite{harris_multi-stage_2021}, \cite{huang_blue_2024}, \cite{cao_autoencoder-based_2024}, \cite{li_sat2scene_2024}, \cite{huang_plantography_2024}, \cite{wang_rip-nerf_2023}, \cite{kant_invs_2023}, \cite{zhou_unsupervised_2022}, \cite{ryu_360_2023}, \cite{sun_f-3dgs_2024}, \cite{zheng_sketch3d_2024}, \cite{vecchio_controlmat_2024}, \cite{tang_realfill_2024}, \cite{xia_survey_2023} \\ 
LPIPS & \cite{sha_nerf-is_2024}, \cite{cao_autoencoder-based_2024}, \cite{li_sat2scene_2024}, \cite{son_singraf_2023}, \cite{kant_invs_2023}, \cite{chen_shaddr_2023}, \cite{ryu_360_2023}, \cite{cheng_cross-modal_2022}, \cite{sun_f-3dgs_2024}, \cite{wu_hyperdreamer_2023}, \cite{vecchio_controlmat_2024},  \cite{tang_realfill_2024}, \cite{xia_survey_2023} \\ 
CLIP Similarity & \cite{dong_coin3d_2024}, \cite{li_icontrol3d_2024}, \cite{alaluf_neural_2023}, \cite{richardson_conceptlab_2024}, \cite{han_scene_2024}, \cite{zhang_clay_2024}, \cite{ryu_360_2023}, \cite{wu_hyperdreamer_2023}, \cite{sella_spicee_2024}, \cite{zheng_sketch3d_2024}, \cite{wang_themestation_2024}, \cite{tang_realfill_2024} \\ 
MSE & \cite{harris_multi-stage_2021}, \cite{huang_blue_2024}, \cite{xia_survey_2023}, \cite{wang_controllable_2024} \\ 
 & \cite{gao_scenehgn_2023}, \cite{alaluf_neural_2023}, \cite{chen_find_2024}, \cite{yiyang_context-aware_2024}, \cite{han_scene_2024}, \cite{wang_elevating_2023}, \cite{eckart_self-supervised_2021}, \cite{sun_f-3dgs_2024}, \cite{nakada_study_2023}, \cite{vecchio_controlmat_2024} \\ 
Accuracy & \cite{ma_multicad_2023}, \cite{wang_elevating_2023}, \cite{eckart_self-supervised_2021}, \cite{de_sousa_ribeiro_object-centric_2024}, \cite{nakada_study_2023}, \cite{wang_embracing_2021}, \cite{li_cad_2024}, \cite{huang_vdam_2022}, \cite{cetinic_understanding_2022}, \cite{ibrahim_explainable_2023} \\ 
Precision & \cite{huang_blue_2024}, \cite{petrov_gem3d_2024}, \cite{nakada_study_2023}, \cite{cetinic_understanding_2022}, \cite{deckers_infinite_2023} \\ 
MMD & \cite{li_sp-gan_2021}, \cite{yiyang_context-aware_2024}, \cite{petrov_gem3d_2024}, \cite{wu_blockfusion_2024}, \cite{nakada_study_2023}, \cite{li_cad_2024} \\ 
IS & \cite{li_icontrol3d_2024}, \cite{jabbar_survey_2021}, \cite{cai_l-magic_2024}, \cite{lin_cascade_2022}, \cite{xia_survey_2023}, \cite{kaswan_generative_2023}, \cite{wang_generative_2021} \\ 
RE & \cite{yang_geolatent_2023}, \cite{yan_online_2021}, \cite{zhang_end--end_2022} \\ 
COV & \cite{li_sp-gan_2021}, \cite{yiyang_context-aware_2024}, \cite{wu_blockfusion_2024} \\ 
1-NNA & \cite{yiyang_context-aware_2024}, \cite{wu_blockfusion_2024}, \cite{kimura_chartpointflow_2021} \\ 
EMD & \cite{gao_scenehgn_2023}, \cite{maesumi_explorable_2023}, \cite{yiyang_context-aware_2024}, \cite{petrov_gem3d_2024}, \cite{zhang_clay_2024}, \cite{kimura_chartpointflow_2021}, \cite{guerrero_matformer_2022} \\ 
\hline
\end{tabular}
\caption{Works classified by metrics used}
\label{tab:metric_references}
\end{table}

To effectively evaluate 3D assets, combining these metrics is essential to address different aspects of authenticity, from geometric accuracy to perceptual quality and distribution realism \cite{xia_survey_2023}. For example, Chamfer Distance can assess point cloud precision, while LPIPS ensures that textures align with human perception. Metrics such as FID and MMD are vital for maintaining diversity and realism in large-scale applications such as gaming or AR/VR asset generation. Furthermore, rendering-based metrics like PSNR and SSIM are particularly useful for evaluating the fidelity of rendered outputs, ensuring visual appeal in real-world applications. This multi-metric approach is crucial for guiding iterative improvements in generative models and advancing their deployment in industries reliant on high-quality 3D asset creation, where maintaining both structural consistency and contextual adaptability is key to creating immersive and functionally coherent virtual worlds while providing a way to evaluate their creativity \cite{franceschelli_creativity_2024}.

\begin{table}[H]

\centering
\renewcommand{\arraystretch}{1.2} 
\setlength{\tabcolsep}{5pt} 
\begin{tabular}{|p{3cm}|p{5cm}|} 
\hline
Type & References \\ 
\hline
Surveys & \cite{de_la_torre_llmr_2024}, \cite{rosenberg_drawtalking_2024}, \cite{polys_prompt_2024}, \cite{lee_knowledge_2023}, \cite{agrawal_image_2024}, \cite{van_der_maden_death_2024}, \cite{tholander_design_2023}, \cite{lee_impact_2024}, \cite{park_we_2024}, \cite{palani_evolving_2024}, \cite{liu_design_2022}, \cite{zhou_eyes_2024}, \cite{sun_predictive_2024}, \cite{sun_generative_2024}, \cite{panchanadikar_im_2024}, \cite{qin_empowering_2023}, \cite{li_learning_2022} \\ 
No Metric & \cite{hertz_spaghetti_2022}, \cite{hu_generative_2024}, \cite{wang_roomdreaming_2024}, \cite{he_creating_2024}, \cite{chen_memovis_2024}, \cite{wang_proteusnerf_2024}, \cite{jiang_synthesizing_2021}, \cite{gottsacker_rlty2rlty_2024}, \cite{merino_interactive_2023}, \cite{dang_worldsmith_2023}, \cite{si_generating_2024}, \cite{jones_shapecoder_2023}, \cite{perla_easi-tex_2024}, \cite{faruqi_style2fab_2023}, \cite{zhong_dreamlcm_2024}, \cite{weng_dream_2024}, \cite{numan_spaceblender_2024}, \cite{chung_promptpaint_2023} \\ 
\hline
\end{tabular}
\caption{References for Papers without Metrics}
\label{tab:no_metric_references}
\end{table}

In table \ref{tab:no_metric_references}, papers were categorized into two different groups that relate to the absence of comprehensive metrics.

\subsection{Domain Application}

Professionals incorporate AI into domains like CAD, AR/VR, and gaming by enhancing workflows, automating complex tasks, and creating scalable content pipelines. For instance, in CAD and industrial design, AI aids in generating diverse ideations \cite{liu_design_2022}, reducing design fixation \cite{liu_3dall-e_2023}, and enabling prompt-based bibliographies \cite{liu_3dall-e_2023}. In gaming and AR/VR, it facilitates high-quality 3D reconstructions and real-time environment generation. These innovations streamline workflows, improve accessibility, and support novel applications like interactive virtual environments and democratized design tools, which were previously limited by technical expertise and resources.
Despite these advancements, specific struggles remain significant. High computational requirements, such as for training diffusion models or processing large datasets, limit real-time applications \cite{hu_cns-edit_2024} and scalability. Training complexity is further exacerbated by dependencies on extensive labeled datasets, such as ShapeNet-v2 or YCB-Video, requiring substantial effort in data collection \cite{zhou_photomat_2023} and preparation. Table \ref{tab:industry_references} compiles papers along with the areas they discuss or directly impact through their proposed innovations.

\begin{table}
\centering
\renewcommand{\arraystretch}{1.2} 
\begin{tabular}{|p{3cm}|p{5cm}|} %
\hline
Industry & References \\ 
\hline
Industrial Design & \cite{liu_3dall-e_2023}, \cite{sha_nerf-is_2024}, \cite{zhang_protodreamer_2024}, \cite{huang_plantography_2024}, \cite{petrov_gem3d_2024}, \cite{jones_shapecoder_2023}, \cite{ma_multicad_2023}, \cite{xiao_text-free_2024}, \cite{li_cad_2024}, \cite{zhou_photomat_2023}, \cite{rios_large_2023}, \cite{upadhyay_floorgan_2023}, \cite{franceschelli_creativity_2024}, \cite{park_we_2024} \\ 
Product Design & \cite{liu_3dall-e_2023}, \cite{hu_cns-edit_2024}, \cite{maesumi_explorable_2023}, \cite{chen_memovis_2024}, \cite{alaluf_neural_2023}, \cite{jones_shapecoder_2023}, \cite{cheng_cross-modal_2022}, \cite{nakada_study_2023}, \cite{guerrero-viu_texsliders_2024}, \cite{li_cad_2024}, \cite{chen_controlstyle_2023}, \cite{zhou_photomat_2023}, \cite{xia_survey_2023}, \cite{son_genquery_2024}, \cite{franceschelli_creativity_2024}, \cite{van_der_maden_death_2024}, \cite{tholander_design_2023}, \cite{lee_impact_2024}, \cite{palani_evolving_2024}, \cite{liu_design_2022}, \cite{sun_generative_2024} \\ 

AR and VR & \cite{sha_nerf-is_2024}, \cite{hu_generative_2024}, \cite{li_sp-gan_2021}, \cite{he_creating_2024}, \cite{de_la_torre_llmr_2024}, \cite{kim_neuralfield-ldm_2023}, \cite{gao_scenehgn_2023}, \cite{maesumi_explorable_2023}, \cite{li_icontrol3d_2024}, \cite{rajaram_blendscape_2024}, \cite{chen_find_2024}, \cite{li_sat2scene_2024}, \cite{zuo_dg3d_2023}, \cite{gothoskar_3dp3_2021}, \cite{yiyang_context-aware_2024}, \cite{gottsacker_rlty2rlty_2024}, \cite{son_singraf_2023}, \cite{wang_rip-nerf_2023}, \cite{petrov_gem3d_2024}, \cite{mohammad_khalid_clip-mesh_2022}, \cite{zhang_vrcopilot_2024}, \cite{si_generating_2024}, \cite{wu_blockfusion_2024}, \cite{cai_l-magic_2024}, \cite{kant_invs_2023}, \cite{rai_assessing_2024}, \cite{chen_shaddr_2023}, \cite{zhang_clay_2024}, \cite{zhou_unsupervised_2022}, \cite{ryu_360_2023}, \cite{eckart_self-supervised_2021}, \cite{perla_easi-tex_2024}, \cite{zhang_3dshape2vecset_2023}, \cite{wu_hyperdreamer_2023}, \cite{zhou_edit3d_2024}, \cite{zhang_text--3d_2023}, \cite{ye_relscene_2024}, \cite{li_imagebind3d_2024}, \cite{zhong_dreamlcm_2024}, \cite{weng_dream_2024}, \cite{huang_placiddreamer_2024}, \cite{yin_text2vrscene_2024}, \cite{numan_spaceblender_2024}, \cite{lee_knowledge_2023},  \cite{bonic_broomrocket_2024}, \cite{zhang_mapa_2024}, \cite{zhang_texpainter_2024}, \cite{wang_themestation_2024}, \cite{agrawal_image_2024}, \cite{yang_diffusion_2023}, \cite{qin_empowering_2023}, \cite{chamola_comprehensive_2024}\\ 
Robotics & \cite{kim_neuralfield-ldm_2023}, \cite{gao_scenehgn_2023}, \cite{gothoskar_3dp3_2021}, \cite{mo_sparse_2024}, \cite{singh_worldgen_2023}, \cite{wu_dreamup3d_2024}, \cite{yan_online_2021}, \cite{eckart_self-supervised_2021}, \cite{agrawal_image_2024}, \cite{li_learning_2022} \\ 
Gaming & \cite{sha_nerf-is_2024}, \cite{li_sp-gan_2021}, \cite{he_creating_2024}, \cite{de_la_torre_llmr_2024}, \cite{kim_neuralfield-ldm_2023}, \cite{hu_cns-edit_2024}, \cite{gao_scenehgn_2023}, \cite{maesumi_explorable_2023}, \cite{li_icontrol3d_2024}, \cite{harris_multi-stage_2021}, \cite{jabbar_survey_2021}, \cite{yang_geolatent_2023}, \cite{li_sat2scene_2024}, \cite{zuo_dg3d_2023}, \cite{gottsacker_rlty2rlty_2024}, \cite{son_singraf_2023}, \cite{wang_rip-nerf_2023}, \cite{merino_interactive_2023}, \cite{awiszus_world-gan_2021}, \cite{petrov_gem3d_2024}, \cite{mohammad_khalid_clip-mesh_2022}, \cite{dang_worldsmith_2023}, \cite{si_generating_2024}, \cite{shen_neural_2024}, \cite{wu_blockfusion_2024}, \cite{cai_l-magic_2024}, \cite{kant_invs_2023}, \cite{singh_worldgen_2023}, \cite{rai_assessing_2024}, \cite{chen_shaddr_2023}, \cite{zhang_clay_2024}, \cite{zhou_unsupervised_2022}, \cite{ryu_360_2023}, \cite{cheng_cross-modal_2022}, \cite{perla_easi-tex_2024}, \cite{zhang_3dshape2vecset_2023}, \cite{zhou_edit3d_2024}, \cite{sella_spicee_2024}, \cite{zhang_text--3d_2023}, \cite{ye_relscene_2024}, \cite{li_imagebind3d_2024}, \cite{zhong_dreamlcm_2024}, \cite{weng_dream_2024}, \cite{huang_placiddreamer_2024}, \cite{zheng_sketch3d_2024}, \cite{vecchio_controlmat_2024}, \cite{hu_diffusion_2024}, \cite{bonic_broomrocket_2024}, \cite{zhang_mapa_2024}, \cite{zhang_texpainter_2024}, \cite{wang_themestation_2024}, \cite{xia_survey_2023}, \cite{wang_generative_2021}, \cite{deckers_infinite_2023}, \cite{palani_evolving_2024}, \cite{zhou_eyes_2024}, \cite{panchanadikar_im_2024}, \cite{qin_empowering_2023}, \cite{chamola_comprehensive_2024}\\ 

Healthcare & \cite{cao_autoencoder-based_2024}, \cite{zhang_end--end_2022}, \cite{wang_embracing_2021}, \cite{wang_controllable_2024}, \cite{kaswan_generative_2023}, \cite{yang_diffusion_2023}, \cite{wang_generative_2021}, \cite{ibrahim_explainable_2023} \\ 
Film-making & \cite{dang_worldsmith_2023}, \cite{shen_neural_2024}, \cite{wu_blockfusion_2024}, \cite{chen_lart_2023}, \cite{zhang_mapa_2024} \\ 
Architecture & \cite{zhao_vaide_2024}, \cite{wang_roomdreaming_2024}, \cite{li_sat2scene_2024}, \cite{jiang_synthesizing_2021}, \cite{huang_plantography_2024}, \cite{han_scene_2024}, \cite{upadhyay_floorgan_2023} \\ 
\hline
\end{tabular}
\caption{Number of References per Industry}
\label{tab:industry_references}
\end{table}

Professionals also face challenges in achieving generalization across diverse scenarios - for example, in robotics or AR/VR, where models struggle with novel views or occlusions, leading to the proposal of innovative frameworks \cite{mo_sparse_2024}. Additionally, integrating AI into existing workflows often proves difficult, as it requires aligning AI outputs with domain-specific standards, such as achieving technical design accuracy in CAD or balancing fidelity and efficiency in 3D reconstruction tasks. These struggles underscore the ongoing need for optimization and tailored solutions.

Designers frequently express frustration over the lack of flexibility and control during the intermediate stages of development when using 3D AI tools \cite{liu_3dall-e_2023}. These systems often generate finalized outputs, whereas designers would prefer the ability to make adjustments and refinements at various stages of the creative process. Table \ref{tab:industry_references} provides information on the number of times each industry was either mentioned by the selected papers or the extent to which the work presented in the papers could be directly translated to practical applications within that industry.

\section{Conclusions}

This review and critical analysis highlight significant advancements in bridging 2D and 3D domains through AI-driven generative models, particularly in scene synthesis and semantic segmentation. Innovations like 3D asset generation from images \cite{liu_3dall-e_2023}, probabilistic scene modeling, and latent tri-plane extrapolation \cite{wu_blockfusion_2024} have enhanced output realism and efficiency both for generation and editing. However, persistent challenges remain, including high computational costs, limited datasets for downstream tasks \cite{kim_neuralfield-ldm_2023}, and difficulties in achieving lifelike detail, which hinder real-time processing and cross-scenario generalization. Rigorous evaluation using geometric metrics (e.g., Chamfer Distance) and perceptual quality metrics (e.g., FID, LPIPS) \cite{xia_survey_2023} is crucial to address these shortcomings and guide future improvements.

The practical application of these technologies in CAD, AR/VR, and gaming underscores their transformative potential, as seen in frameworks like CLAY \cite{zhang_clay_2024} and Edify3D \cite{nvidia2024edify3dscalablehighquality}. Scalability and industry adoption face hurdles due to training complexities, resource-intensive workflows \cite{hu_cns-edit_2024}, and the scarcity of diverse, high-quality datasets \cite{wang_roomdreaming_2024}. While the survey maps the current landscape and identifies critical innovation areas, overcoming these barriers requires optimized data strategies, architectural refinements (e.g., diffusion transformers \cite{zhang_clay_2024}), and techniques like proxy-guided conditioning \cite{dong_coin3d_2024} to balance efficiency with output fidelity.

Looking ahead, advancing 3D content generation demands enhanced model architectures (e.g., Gaussian splatting \cite{sun_f-3dgs_2024}), expanded multimodal input integration \cite{zhou_edit3d_2024}, and robust curated datasets. Addressing limitations such as computational bottlenecks in diffusion models and generalization issues in GANs \cite{son_singraf_2023} will foster adaptable, scalable systems capable of producing immersive environments that meet creative and industrial demands.

\printbibliography

%





\ifCLASSOPTIONcaptionsoff
  \newpage
\fi








\vfill


\end{document}